\documentclass[sigconf]{acmart}
\AtBeginDocument{%
  }

\copyrightyear{2025}
\acmYear{2025}
\setcopyright{rightsretained}
\acmConference[UMAP '25]{33rd ACM Conference on User Modeling, Adaptation and Personalization}{June 16--19, 2025}{New York City, NY, USA}
\acmBooktitle{33rd ACM Conference on User Modeling, Adaptation and Personalization (UMAP '25), June 16--19, 2025, New York City, NY, USA}
\acmDOI{10.1145/3699682.3730972}
\acmISBN{979-8-4007-1313-2/2025/06}

\begin{document}

\title{Predicting Movie Hits Before They Happen with LLMs}

 \author{Shaghayegh Agah}
\authornote{These authors contributed equally and share first-author status.}
\affiliation{%
  \institution{Comcast Technology AI}
  \city{Sunnyvale CA}
  \country{USA}}
\email{shaghayegh_agah@comcast.com}

\author{Yejin Kim}
\authornotemark[1]
\affiliation{%
  \institution{Comcast Technology AI \& \\George Washington University}
  \city{Washington DC}
  \country{USA}}
\email{yejin_kim@comcast.com}

\author{Neeraj Sharma}
\authornotemark[1]
\affiliation{%
  \institution{Comcast Technology AI}
  \city{Sunnyvale CA}
  \country{USA}}
\email{neeraj_sharma@comcast.com}

\author{Mayur Nankani}
\authornotemark[1]
\affiliation{%
  \institution{Comcast Technology AI}
  \city{Sunnyvale CA}
  \country{USA}}
\email{mayur_nankani@comcast.com}

\author{Kevin Foley}
\affiliation{%
  \institution{Comcast Technology AI}
  \city{Washington DC}
  \country{USA}}
\email{kevin_foley@comcast.com}

\author{H. Howie Huang}
\authornote{Corresponding author.}
\affiliation{%
  \institution{George Washington University}
  \city{Washington DC}
  \country{USA}}
\email{howie@gwu.edu}

\author{Sardar Hamidian}
\authornotemark[1]
\authornotemark[2]
\affiliation{%
  \institution{Comcast Technology AI}
  \city{Washington DC}
  \country{USA}}
\email{sardar_hamidian@comcast.com}


\renewcommand{\shortauthors}{Agah et al.}

\begin{abstract}
Addressing the cold-start issue in content recommendation remains a critical ongoing challenge. In this work, we focus on tackling the cold-start problem for movies on a large entertainment platform. Our primary goal is to forecast the popularity of cold-start movies using Large Language Models (LLMs) leveraging movie metadata. This method could be integrated into retrieval systems within the personalization pipeline or could be adopted as a tool for editorial teams to ensure fair promotion of potentially overlooked movies that may be missed by traditional or algorithmic solutions. Our study validates the effectiveness of this approach compared to established baselines and those we developed.
\end{abstract}






\maketitle
\section{Introduction}
Listwise ranking approaches have been highly effective in personalization tasks \cite{pang2020setrank,zhuang2023rankt5,pang2017deeprank}. Hybrid collaborative-content approaches have further advanced the field by combining user behavior encoding with content features and leveraging contextual and temporal signals to build state-of-the-art systems \cite{chen2019joint,cheng2016wide,zheng2017joint}. Despite these advances, promoting new contents and addressing the cold start problem remains a significant challenge, especially in industries where content evolves rapidly \cite{lu2020meta,wei2017collaborative}. In the entertainment domain, many companies rely on in-house editorial teams with domain expertise to give new content a chance. At the same time, automated methods for predicting popularity often suffer from biases such as seasonality or past popular programs. As a result, many newly released content may remain underexposed, negatively
impacting user-item diversity and engagement. The rapid increase in content across platforms and languages makes relying on human editors or existing machine learning (ML) for new content promotion difficult and unscalable.

To address this challenge, we investigate using LLMs as a complementary
tool to traditional editorial and ML processes. Our approach employs LLMs to assess newly released content based on metadata as a prompt, generating not only a probability estimate of the hit potential of a content piece but also a transparent reasoning for its prediction. While many external factors beyond a movie's quality influence its future popularity, we hypothesize that content-specific attributes, combined with intrinsic knowledge of LLMs, can reasonably assess a film's likelihood of becoming popular. Our results indicate that LLMs can identify potential hit movies well before they gain significant audience attention. This approach could be used as an aided system for the editorial team or an automated retrieval solution in content recommendation
pipeline.

\section{Methodology}
Our methodology involves four steps: (1) building a proper dataset (2) establishing a baseline for comparison, (3) evaluating LLM effectiveness in natural language reasoning vs. latent embedding space, and (4) optimizing performance using LLMs in generative mode and prompt engineering to predict a movie's hit potential. We conducted experiments using Llama 3.3 and 3.1 \cite{dubey2024llama}, applying various prompts to analyze their predictive capabilities.

\subsection{Dataset}
To the best of our knowledge, no publicly available dataset in the
movie domain allows for a direct evaluation of our hypothesis.
Most existing datasets were built before the rise of LLMs and lack
comprehensive metadata, making it challenging to construct an un-
biased experimental setup. To address this, we designed a benchmark dataset from our platform of newly released and watched items and tracked their progression to popularity over time. Our entertainment platform serves tens of millions of users, who engage with movies and TV content either directly or through third-party apps. We focused on movies because they are less influenced by external knowledge than TV series, improving the reliability of experiments. 
 Movie popularity is defined by user interactions. The dataset consists of newly released movies as input and movies that later became popular as labels. To find the labels, we analyzed how long it takes for a movie to become popular, examining various time windows and popularity list sizes. We evaluated LLM and baseline model performances across time-based settings. Metadata such as genre, synopsis, content ratings, era, cast, crew, mood, awards, and character types are provided to the LLM as input. For comparison, we set two baselines, 1) Random and 2) Popular Embedding (PE) described in section \ref{baseline}.

\subsection{LLM-based Popularity}

We use LLMs as the primary ranking mechanism to predict content popularity, generating ranked lists with popularity scores and reasoning while exploring different prompt engineering strategies. Our prompts guide the LLM to predict a movie's popularity based on available metadata. 


 \subsubsection{Prompt engineering}
Our prompting strategy is designed to guide the LLM in predicting movie popularity based on available metadata. The prompts frame the model as a movie popularity prediction expert, assisting an editor in selecting upcoming popular titles. The model is instructed to reorder the provided list strictly without adding new items and to return results in a structured JSON format. The response includes a ranked list of movies, assigned popularity scores, reasoning behind the rankings, and awareness of prior data points used. By incorporating multiple levels of metadata, our approach enhances the model's contextual understanding, improving its ability to forecast emerging trends in entertainment.



\section{Experimental Conditions}
The experiment proceeded mainly in two steps: 1) Establishing a baseline by evaluating multiple variants and selecting the one that outperformed the random-ordering model. 2) Evaluating LLMs in generative mode against the selected baseline to increase the challenge. There is a margin of 6-12 months between LLMs' knowledge cutoff and movie release dates. 

\paragraph{\textbf{Baseline Evaluation}}
\label{baseline}
We created a baseline model by generating semantic representations of program metadata using embeddings from BERT \cite{devlin2019bert}, Linq-Embed-Mistral (7B) \cite{choi2024linq} and Llama 3.3 (70B) \cite{dubey2024llama}. We selected Linq-Embed-Mistral for its top-rank on the MTEB leaderboard \cite{mteb_leaderboard}. The 70B model was quantized to 8 bits due to the experiment environment. The predicted order of popularity is determined by the cosine similarity of the program embeddings with the average embedding of the top-\textit{100} popular items in the weeks leading up to the release date. To evaluate the performance of each approach, we compared the improvement of our main metrics against a random ordering of the candidate list. Table \ref{pe} shows that BERT V4 and Linq (7B) V4 achieve the most significant gains in top-\textit{3} ranking performance, although they slightly underperformed in predicting the most popular item. Since the overall benefits outweighed the drawbacks and BERT is a lightweight model, we chose it as the baseline for comparing LLM performance in the next section.
\begin{table}[ht]
\centering
\caption{PE Models Performance Improvement (\%) vs Random. Metadata (MD) explanations: "V1": Genre, "V2": Synopsis, "V3": V1 + V2 + Content ratings, Character types, Mood, Era, "V4": V3 + Cast, Crew, Awards}  
\begin{tabular}{lccccc}
\hline
\textbf{Model}  & \textbf{MD} & \textbf{ACC@1} & \textbf{RR} & \textbf{NDCG@3} & \textbf{RC@3}\\ \hline
BERT & V2 & 166   & 58.02  & 43.24 & 28.57   \\ 
BERT & V4 & 100   & 39.33  & 55.73 & 47.62 \\ 
Linq-7B & V4   & 100  & 39.39  & 55.78 & 47.62 \\ 
Llama-70B & V4   & 33.33  & 23.78  & 9.47 & 7.01 \\ 
Llama-70B & V3   & 33.33 & 11.70  & 27.8 & 19.05 \\ \hline
\end{tabular}
\label{pe}
\end{table}
\paragraph{\textbf{LLM Evaluation}} We evaluated performance using pairwise and listwise ranking strategies \cite{qin2023large,zhuang2024setwise} and conducted ablation studies to assess the impact of different metadata attributes on model performance. Using full, non-quantized models, this setup enabled a systematic analysis of the effectiveness of LLM-based predictions compared to metadata-driven embeddings while ensuring a practical and reproducible evaluation.
\begin{table}[ht]
\setlength{\abovecaptionskip}{6pt}
\setlength{\belowcaptionskip}{6pt}

\centering
\caption{LLM Performance Improvement (\%) vs BERT V4}
\begin{tabular}{lccccc}
\hline
\textbf{Model}  & \textbf{MD} & \textbf{ACC@1} & \textbf{RR} & \textbf{NDCG@3} & \textbf{RC@3}\\ \hline
Llama-405B & V1 & -80.00 & -24.42 & -17.42 & -16.71  \\
Llama-405B  & V2 & -25.00 & -7.17 & 3.39 &19.07 \\
Llama-405B  & V3 & -16.67 & 5.66 & 0.99 & 7.98\\
Llama-405B  & V4 & \textbf{28.33} & \textbf{22.46} & \textbf{12.90} & \textbf{31.42}  \\
Llama-70B  & V1 & -65.00 & -11.92 & -9.03 & -10.35  \\
Llama-70B  & V2 & -8.33 & 0.05 & 5.69 & 17.11  \\
Llama-70B  & V3 & -42.00 & -20.89 & -16.29 & -18.86 \\ 
Llama-70B  & V4 & -6.67 & 7.36 & \textbf{8.45} & \textbf{25.03}  \\
Llama-8B  & V1 & -82.50 & -26.86 & -15.74 & -16.15  \\
Llama-8B  & V2 &  \textbf{1.33} & \textbf{8.88} & -5.76 & -3.01  \\ 
Llama-8B  & V3 & -66.17 & -13.85 & -14.89 & -12.54  \\
Llama-8B  & V4 & -84.26 & -38.79 & 2.01 &-2.94  \\ 
\hline
\end{tabular}
\label{table:model_performance}
\end{table}

\subsection{Metrics}
To evaluate the effectiveness of LLMs in predicting movie popularity, we use both ranking-based and classification-based metrics, focusing on accurately placing the top-3 most popular contents. Specifically, we assess: 1) Accuracy@\textit{1} (ACC@\textit{1}): how often the most popular item is correctly predicted in the first position. 2) Reciprocal Rank (RR): represents the reciprocal rank of the top popular item in the predicted list. 3) Normalized Discounted Cumulative Gain (NDCG@\textit{k}): measures how the ranking predictions are aligned with actual popularity. The relevance score of each item is determined based on its popularity score. 4) Recall@\textit{3}: percentage of the top 3 most popular items in the top-\textit{3} predicted items. Given that most interactions occur within the top indices of the menu, we restrict our evaluation to lower @\textit{k}  values for a more comprehensive performance assessment.

\subsection{Results and Evaluations}

We evaluated the performance of different versions of Llama models (3.1 (8B), 3.1 (405B), 3.3 (70B)) by measuring their respective metric improvements relative to the baseline.  A series of prompt variations, from basic to information-rich, were used to test each model. The results are presented in Table \ref{table:model_performance}, with the highest and second highest values in bold. For each variation of the model and prompt, we performed 10 experiments, and the reported results reflect the average percentage improvement across all analyses. 
The best performance is achieved when using Llama 3.1 (405B) with the most informative prompt, followed by Llama 3.3 (70B). Based on the observed trend, when using a complex and lengthy prompt (MD V4), a more complex language model generally leads to improved performance across various metrics. However, it is sensitive to the type of information added. As demonstrated, adding the top 5 cast awards to the prompt (MD V4) significantly outperformed a prompt without them (MD V3). This is observed for the larger model (405 \& 70B). For the smaller Llama 3.1 (8B) model, performance is increased with slightly more complex prompts from genre to synopsis, but further complexity decreased the performance. This could be attributed to its smaller size, which limits its ability to handle complex prompts. The model likely struggles to generalize effectively, leading to suboptimal predictions. Across all model prompts with genres only (MD V1), performance below the baseline indicates insufficient information for the LLM to generate meaningful conclusions.
In addition, we conducted statistical analysis for top performing models, and the average results fall within the expected statistical range, indicating consistency and reliability.

\subsection{Pairwise vs Listwise}

Previous studies showed improved document ranking given a query using pairwise ranking vs listwise \cite{qin2023large,zhuang2024setwise}. For our work, pairwise ranking did not improve key metrics and sometimes underperformed compared to listwise. Document ranking based on a query that has clear intent, and candidates can be ranked using relevancy scores is different than the broad open-ended query tested in this work. In our scenario — ranking newly released movies — there is no inherent connection between the candidate movies. A random selection of new releases may include films from different genres, directors, and styles with no common query binding them. This, along with fewer items per list, makes pairwise comparisons for our context less effective. 

\subsection{Insights from LLM Experiments}
We conducted experiments to evaluate LLMs' performance by testing various prompts, altering movie list orders, and exploring different timelines to assess the stability of results. Key insights include the importance of refining the parsing process. Initial attempts to extract structured outputs using regex were inconsistent, so we adjusted prompts to instruct the LLM to generate responses in a predefined JSON format and added negative instructions. This significantly improved the responses' reliability. We also observed variations across LLMs over identical prompts, emphasizing the need for model-specific adjustments and adaptive strategies. Another challenge was inconsistent response lengths, addressed through post-processing to filter mismatched responses. We also found that a single structured input was more effective than using multiple lists with positional dependencies.
\section{Conclusion and Future Work}
Predicting whether a movie will be a hit or a miss is challenging, especially with the constant release of new movies. The industry still relies on subjective, non-scalable methods to explore and promote content. In this work, we propose leveraging LLMs to tackle this challenge. We evaluate LLMs' performance against several baselines, using a dataset from actual production data, and examine their performance across different data and experimental settings.

Our results show that LLMs, when using movie metadata, can significantly outperform the baselines. This approach could serve as an assisted system for multiple use cases, enabling the automatic scoring of large volumes of new content released daily and weekly. By providing early insights before editorial teams or algorithms have accumulated sufficient interaction data, LLMs can streamline the content review process. With continuous improvements in LLM efficiency and the rise of recommendation agents, the insights from this work are valuable and adaptable to a wide range of domains.


\bibliographystyle{ACM-Reference-Format}
\bibliography{sample-base}


\end{document}